\documentclass[12pt]{article}
\usepackage{amsmath,amsfonts,amssymb,amsthm,amstext,amscd,eucal}
\usepackage[all]{xy}

\makeatletter \@addtoreset{equation}{section}

\makeatletter\renewcommand\section{\@startsection {section}{1}{\z@}%
                                   {-3.5ex \@plus -1ex \@minus -.2ex}
                                   {2.3ex \@plus.2ex}%
                                   {\normalfont\large\bfseries}}
\renewcommand\subsection{\@startsection{subsection}{2}{\z@}%
                                     {-3.25ex\@plus -1ex \@minus -.2ex}%
                                     {1.5ex \@plus .2ex}%
                                     {\normalfont\bfseries}}

\newcommand{\be}{\begin{equation}}
\newcommand{\ee}{\end{equation}}
\newcommand{\bse}{\begin{subequations}}
\newcommand{\ese}{\end{subequations}}
\newcommand{\beq}{\begin{eqnarray}}
\newcommand{\eeq}{\end{eqnarray}}
\newcommand{\bea}{\begin{eqnarray}}
\newcommand{\eea}{\end{eqnarray}}

\def\[{\left [}
\def\]{\right ]}
\def\({\left (}
\def\){\right )}

\def\r2{\sqrt{2}}

\newcommand{\cf}{{\em cf.}\ }

\def\sst#1{{\scriptscriptstyle #1}}

\def\1{{\sst{(1)}}}

\newcommand{\ads}[1]{${\rm AdS}_{#1}$}

\newcommand{\bbibitem}[1]{\bibitem{#1}\marginpar{#1}}

\def\Label#1{\label{#1}%
  \smash{\hbox to0pt{\raise1ex\hbox{\tiny[#1]}\hss}}}
\def\noLabels{\let\Label=\label}
\def\nobbibitem{\let\bbibitem=\bibitem}

\begin{document}

\begin{titlepage}

\begin{flushright}\vspace{-2cm}
{\small
{\tt arXiv:1011.1879} \\
IPM/P-2010/045
 }\end{flushright} \vspace{5 mm}

\begin{center}
\centerline{{\Large{\bf{Near Horizon Limits of Massless BTZ  and Their CFT Duals}}}} \vspace{4mm}

{\large{{\bf
Jan de Boer\footnote{e-mail:
J.deBoer@uva.nl}$^{,ab}$,  M.M. Sheikh-Jabbari\footnote{e-mail:
jabbari@theory.ipm.ac.ir}$^{,c}$
and Joan Sim\'on\footnote{e-mail: j.simon@ed.ac.uk}$^{,d}$ }}}
\\

\vspace{5mm}

\bigskip\medskip
\begin{center}
{$^a$ \it Instituut voor Theoretische Fysica, Valckenierstraat 65,\\
1018XE Amsterdam, The Netherlands}\\
\smallskip
{$^b$ \it Gravitation and AstroParticle Physics Amsterdam}\\
\smallskip
{$^c$ \it School of Physics, Institute for Research in Fundamental
Sciences (IPM),\\ P.O.Box 19395-5531, Tehran, Iran}\\
\smallskip
{$^d$ \it School of Mathematics and Maxwell Institute for Mathematical Sciences,\\
King's Buildings, Edinburgh EH9 3JZ, United Kingdom}\\
\smallskip
\end{center}
\vfil

\end{center}
\setcounter{footnote}{0}

\begin{abstract}
\noindent We consider the massless BTZ black hole and show that it
is possible to take its ``near horizon'' limit in two distinct ways.
The first one leads to a null self-dual orbifold of AdS${}_3$ and the second
to a singular space-like AdS${}_3/\mathbb{Z}_K$ orbifold in
the large $K$ limit, the ``pinching orbifold''. We show that from the dual 2d CFT viewpoint, the
null orbifold corresponds to the $p^+=0$ sector of the Discrete
Light-Cone Quantization (DLCQ) of the 2d CFT where a chiral sector
of the CFT is decoupled, while the pinching orbifold
corresponds to taking an infinite mass gap limit in both the right and
left sectors of the 2d CFT, essentially leaving us with the states
$L_0=\bar L_0=\frac{c}{24}$ only. In the latter case, one can
combine the near horizon limit with sending the 3d
Planck length $\ell_P$ to zero, or equivalently the dual 2d CFT central charge $c$ to infinity. We provide
preliminary evidence that in that case some nontrivial dynamics may survive the limit.

\end{abstract}
\vspace{0.5in}

\end{titlepage}
\renewcommand{\baselinestretch}{1.05}  
\tableofcontents


\section{Introduction}
\pagestyle{plain}
\baselineskip=19pt

Despite the progress made in understanding the statistical mechanical origin of black hole thermodynamics, the identification
of the underlying microscopic degrees of freedom for generic black holes remains an open problem.
In most examples
in which such an identification was achieved, the black hole possesses an AdS${}_3$ throat in its
near-horizon limit, and the microscopic degrees of freedom are captured by a two-dimensional
conformal field theory. In some other cases, the near horizon geometries resemble but are not quite AdS${}_3$, and it is clearly important to understand the precise meaning of such geometries, and their connection to the degrees of freedom of two-dimensional field theories. Motivated by this, we decided to analyse and interpret the various possible
near-horizon limits that one can take in the case of the BTZ black hole \cite{Banados:1992wn}.

In \cite{Balasubramanian:2009bg} this analysis was initiated by considering extremal BTZ black holes.
In particular it was noted that (i) the spacelike circle at the causal boundary of the extremal BTZ geometry becomes
a light-like circle at the horizon and, (ii) moving from the boundary to the horizon is exactly like boosting
the spacelike circle so that it becomes approximately null. Next, it was noted that the dual 2d CFT resides on an
$\mathbb{R}\times S^1$ boundary, and that this boosting to the speed of light exactly matches Seiberg's
definition \cite{Seiberg} of the Discrete Light-Cone Quantization (DLCQ) of the 2d CFT. Since
the eigenvalues of left and right excitation operators $L_0-c/24$ and $\bar L_0-c/24$ are scaled with opposite
boost factors (if one of them is scaled up by the factor $\gamma$ the other is scaled by $\gamma^{-1}$),
the boosting will create an infinite mass gap in one of the sectors, say the $\bar L_0$ sector, if we
intend to keep the mass scale of the $L_0$ sector finite. That is, in the DLCQ description, due to the
infinite mass gap, we cannot excite the $\bar L_0$ sector and it is to be set to its ground state
with $\bar L_0=\frac{c}{24}$. On the other hand, the $L_0$ sector which has a finite mass gap can be
arbitrarily excited. In other words, in the DLCQ description a non-chiral 2d CFT reduces to a chiral sector.

Since the DLCQ prescription emerges when approaching the horizon, the near-horizon geometry of the
extremal BTZ black hole should be the holographic dual of the DLCQ of the 2d CFT. As shown in
\cite{Balasubramanian:2009bg}, the near-horizon geometry of extremal BTZ is the so-called spacelike
self-dual \ads3 orbifold \cite{selfdual}. It contains an AdS${}_2$ factor which cannot be excited and which is the
geometric manifestation of the frozen $\bar L_0$ sector. The excitations in the dynamical $L_0$
sector crucially involve the remaining third space-time coordinate.

In \cite{Kerr-CFT} it was proposed that chiral 2d CFT's may provide
us with the ``dual CFT'' description of near horizon extremal 4d Kerr black hole, and this idea has subsequently been extended to various other types of near-extremal black holes. This
proposal was given preliminary backing through an asymptotic symmetry group analysis and
a successful computation of the black hole entropy using Cardy's formula.
However, it is not clear if this chiral
CFT captures any of the dynamics around an extremal black hole \cite{No-dynamics}.
In particular, for
the simplest extremal black hole, i.e. extremal BTZ, in \cite{Balasubramanian:2009bg} we
showed that the dual description of the self-dual \ads3 geometry that appears
in the near horizon limit is indeed a
thermal state in a chiral 2d CFT with temperature determined by the light-cone
momentum $p^+$ of the dual DLCQ CFT. Consistently, the light-cone momentum $p^+$ is also the
only free parameter of the self-dual \ads3 geometry. Because the right-movers are frozen
very little dynamics is left in this case.

One issue which was left open in the above analysis was the
geometry representing the ``ground state'', i.e. the $p^+=0$ sector, of the DLCQ chiral CFT.
Naively, this geometry should be obtainable from a suitable near-horizon
limit of massless BTZ, and as we show below this is indeed the case. We will
show that it is the null self-dual AdS${}_3$ orbifold whose metric is
given in (\ref{null-orbifold}) which is dual to
the ground state of the chiral CFT, whereas the spacelike self-dual AdS${}_3$ orbifold (\ref{eq:c1})
corresponds to a generic state with $p^+\neq 0$ \cite{Balasubramanian:2009bg}.

The null self-dual orbifold was recently found in a near-horizon limit in \cite{compere},
but is not the only geometry which has appeared in the
literature in the study of near-horizon limits of extremal vanishing horizon
black holes. A different geometry, which we will refer to as the pinching
orbifold of \ads3, has also appeared several times starting with the work of
\cite{Bardeen-Horowitz}, and more recently in \cite{FGMS, guica,terashima,Matsuo-Nishikoa}.
Its geometry is exactly the same as that of $M=0$ BTZ, except that the periodicity of the
asymptotic $S^1$ is changed from $2\pi$ to $2\pi\epsilon$, the near-horizon limit being the $\epsilon\to 0$
limit. We will review the fairly straightforward near-horizon limit of
the $M=0$ BTZ black hole which produces this pinching orbifold and show
that in this limit both sectors of the 2d CFT are frozen rather than just one.

Since the massless BTZ black hole has a naked singularity, it is not unexpected that both the null self-dual orbifold, as well as the pinching orbifold, are not good reliable gravitational backgrounds. Nevertheless, we hope our observations help to
clarify the physical interpretation of the appearance of these two distinct singular geometries arising in different
near horizon limits.

In the case of the pinching orbifold, as one goes to lower and lower
energies, both the left and right moving sectors lose their dynamics
because the dual CFT has a mass gap. For CFT's with an \ads3 dual this mass
gap is typically of the order of $1/c$. One could therefore expect
that one might be able to retain some dynamics by combining a near-horizon
limit with a $c\to \infty$ limit. We will present a preliminary
analysis of this possibility in section~4, after discussing the various
near-horizon limits in section~2 and their dual interpretation in
section~3. We conclude with some discussion of our results and
provide a comprehensive review of near-horizon limits of BTZ geometries
in the appendix.

In \cite{EVH-us}, we will discuss various generalisations of all these
near-horizon limits to other black holes, in particular to two R-charge
AdS${}_5$ black holes, following up on previous work in
\cite{FGMS, Balasubramanian:2007bs}.

\paragraph{Note added:} While developing these ideas, we became aware of the work \cite{vijay-simon} which has some overlap with our discussion regarding the null self-dual orbifold and its interpretation as the $p^+=0$ sector of the DLCQ of the 2d CFT.

\section{Near horizon (low energy) limits of massless BTZ}\label{NH-limits-section}

Near horizon limits can generically be interpreted as low energy limits. There may exist different inequivalent low energy limits to be considered in a given physical system, and consequently, one may expect the existence of more than a single near horizon limit when realising these notions holographically. If so, these different limits may focus on different physics of the same theory.

We will study this possibility for the massless BTZ black hole. Since this has a naked singularity, one expects its near horizon geometry to be singular too. Keeping the radius of AdS${}_3$ $\ell_3$ fixed, and remembering there are no bulk degrees of freedom, the resulting geometry must be a quotient of AdS${}_3$.\footnote{The set of inequivalent abelian quotients of AdS${}_3$ appears already in \cite{henneaux-btz}. A complete analysis on d=3 self-dual quotients along with their supersymmetry appears in \cite{selfdual}. This was extended to higher dimensions in
\cite{Madden:2004yc,FigueroaO'Farrill:2004yd}, including a complete supersymmetry analysis \cite{FigueroaO'Farrill:2004yd} and geometrical properties of the causally well-behaved ones in \cite{FigueroaO'Farrill:2004bz}, where it was already mentioned that the null self-dual orbifold should play a role in any DLCQ formulation of the dual field theory.}

In the following we partially review some set of well known relations between massless BTZ black holes, extremal BTZ black holes and their near horizon limits that will allow us to get the appropriate physical intuition to correctly interpret the two different inequivalent near horizon limits that exist for massless BTZ, one leading to the {\it null self-dual} orbifold and the second to a singular geometry that we shall refer to as a {\it pinching orbifold}.

\subsection{Null self-dual \ads3 orbifold as the near horizon limit of massless
BTZ}\label{NH-Null-orbifold}

Consider an extremal BTZ black hole
\begin{equation}
  ds^2 = -\frac{(r^2-r_h^2)^2}{r^2\,\ell_3^2}\,dt^2 + \frac{\ell_3^2\,r^2}{(r^2-r_h^2)^2}\,dr^2 + r^2\,\left(d\phi - \frac{r_h^2}{r^2}\,\frac{dt}{\ell_3}\right)^2\,,
\label{eq:btz}
\end{equation}
and take the near horizon limit
\begin{equation}
  r^2 = r_h^2 + \epsilon\rho\,, \quad \quad t = \frac{\tau}{\epsilon}\,, \quad \quad \varphi = \phi - \frac{\tau}{\epsilon\,\ell_3} \quad \quad \epsilon\to 0\,.
\end{equation}
This gives rise to the spacelike self-dual orbifold \cite{Balasubramanian:2009bg} \footnote{The importance of this geometry for the physics of extremal black holes was already emphasized some time ago in \cite{andyads2,asad}.} %
\begin{equation}
\begin{split}
  ds^2 &= \frac{\ell_3^2}{4}\frac{d\rho^2}{\rho^2} - \frac{\rho^2}{r_h^2}\frac{d\tau^2}{\ell_3^2} + r_h^2\left(d\varphi + \frac{\rho}{r_h^2}\,\frac{d\tau}{\ell_3}\right)^2\\
  &= \frac{\ell_3^2}{4}\frac{d\rho^2}{\rho^2}+ 2\frac{\rho}{\ell_3} d\tau d\varphi+ r_h^2 d\varphi^2
  \,.
\end{split}\label{eq:c1}
\end{equation}
Its causal boundary is a null cylinder parameterised by the null non-compact $\tau$
direction and the null circle in the $\varphi$ direction. The $\varphi$ direction at generic $\rho$ is a spacelike
circle, becoming null at the boundary, while $\tau$ is always null.
It was discussed in \cite{Balasubramanian:2009bg} that the self-dual
\ads3 orbifold geometry is dual to a chiral CFT which is what remains of a
generic 2d CFT in the DLCQ description; this chiral CFT
resides on the null cylinder causal boundary of the self-dual
orbifold. Once the periodicity of the $\varphi$ direction is fixed
to $2\pi$, the only parameter of the self-dual orbifold metric,
$r_h$, is then related to the value of the light-cone momentum $p^+$
defining the DLCQ sector:
\be\label{LC-momentum}%
p^+=\frac c6 \left(\frac{r_h}{\ell_3}\right)^2\,.
\ee
 To obtain \eqref{LC-momentum}
we have used the fact that $p^+=L_0-\frac{c}{24}$ and that the
mass $M$ or angular momentum $J$ of the original extremal BTZ and $L_0$ are related as\footnote{In our conventions
the ADM mass and angular momentum of a generic BTZ geometry with inner and outer horizons $r_\pm$ are given by%
\be\label{mass-J}%
M\ell_3=\frac{r_+^2+r_-^2}{8G_3\ell_3}\ ,\quad J=\frac{r_+r_-}{4G_3\ell_3}\,.%
\ee%
}%
\be M\ell_3=J= p^+=\frac{r_h^2}{4\ell_3 G_3}\,,\qquad
c=\frac{3\ell_3}{2G_3}\,. %
\ee %
As discussed in \cite{Balasubramanian:2009bg} the self-dual orbifold
is a thermal state in the DLCQ of the 2d CFT at temperature\footnote{The left and right temperatures
are given for a generic BTZ is given by%
\be\label{TL-TR}
T_L=\frac{r_++r_-}{2\pi\ell_3} \,,\qquad T_R=\frac{r_+-r_-}{2\pi\ell_3}
\ee
and the Hawking temperature of the BTZ black hole $T_H$ is $\frac{2}{T_H}=\frac1{T_L}+\frac1{T_R}$\,.}%
\be\label{T-self-dual}%
T_{\rm{self-dual}}=\frac{r_h}{\pi \ell_3}=\sqrt{\frac{6p^+}{\pi^2 c}}\,. %
\ee%

The $r_h\to 0$ limit of the spacelike self-dual orbifold sends the temperature \eqref{T-self-dual} to zero and yields the metric
\be\label{null-orbifold}%
ds^2=r^2 dx^+dx^-+\ell_3^2\frac{dr^2}{r^2}\,,\qquad x^-\sim
x^-+2\pi\,, \ee%
where we have conveniently renamed $\rho=r^2$, $\varphi$ as $x^-$
and $\tau=2\ell_3 x^+$. Notice the causal character of the compact direction $x^-$ has changed, from an everywhere spacelike direction (except at the boundary) to an everywhere null direction. By construction, this corresponds to the null self-dual orbifold. The latter has the same boundary as \eqref{eq:c1}, and so it could be viewed as belonging to the same semiclassical Hilbert space that one could construct through a Brown-Henneaux type analysis \cite{brown-henneaux}. Thus, the null orbifold should correspond to a state in the DLCQ CFT. Since the $r_h\to 0$ limit corresponds to $p^+\to 0$, it is natural to relate the null orbifold to the $p^+=0$ sector of the DLCQ CFT, which is the most natural identification to make especially when working in some Poincar\'e patch description of AdS${}_3$.

Instead of taking the vanishing horizon limit after the near horizon limit, we can study the near horizon limit of the massless BTZ black hole. This singular black hole is
\begin{equation}
ds^2 = r^2d\tilde x^+d\tilde x^- +\ell^2_3 \frac{dr^2}{r^2}\quad \quad
\tilde x^\pm = \phi \pm t\,, \qquad \phi\sim \phi+2\pi\,.
 \label{eq:massless}
\end{equation}
Consider the near horizon limit%
\be%
r=\epsilon \rho\,,\qquad \tilde x^-=x^-\,,\qquad \tilde
x^+=\frac{x^+}{\epsilon^2}\,,\qquad \epsilon\to 0\,. %
\label{eq:limit1}
\ee%
The lightlike direction $\tilde x^+$ effectively decompactifies, for the same reason as for the spacelike self-dual orbifold, while $x^-$ remains compact $x^-\sim x^-+2\pi$. Thus, the near horizon limit \eqref{eq:limit1} of a massless BTZ black hole is the null self-dual \ads3 orbifold \eqref{null-orbifold}.

\paragraph{Exciting the null self-dual orbifold :} If this proposal is correct, the spacelike self-dual orbifold \eqref{eq:c1} should be viewed as an excitation over the null orbifold%
\begin{equation}%
ds^2 = \frac{\ell_3^2}{z^2}\left(dx^+dx^- + dz^2\right)\,, \quad
\quad x^-\sim x^- + 2\pi\,,
 \label{eq:null}
\end{equation}
In particular, injecting some chiral momentum into the system keeping its causal null cylinder boundary should correspond to the spacelike self-dual orbifold. This is achieved by adding some wave to the conformally flat metric
\begin{equation}
ds^2 = \frac{\ell_3^2}{z^2}\left[dx^+dx^- + kz^2(dx^-)^2+
dz^2\right]\,,
\end{equation}
Since there are no propagating degrees of freedom in d=3, the latter is locally AdS${}_3$, and it is indeed isometric to the spacelike self-dual orbifold \eqref{eq:c1}, with $r^2_h$ being replaced with $k\ell_3^2$. Recalling \eqref{LC-momentum}, this corresponds to a DLCQ sector with
light-cone momentum%
\be%
p^+=\frac c6 k \,. %
\ee%
This observation is consistent with the interpretation given above based on \cite{Balasubramanian:2009bg}. Indeed, among the Brown-Henneaux diffeomorphisms, there is only one chiral sector that preserves the null boundary structure shared by both orbifolds. The spacelike self-dual one is obtained after the finite action of one of these diffeomorphisms belonging to this same chiral sector.

All these observations are consistent with the well-known fact that
extremal BTZ is a chiral excitation above the massless BTZ black
hole \cite{Cvetic:1998jf,Brecher:2000pa}. To see this, consider the massless BTZ black
hole \eqref{eq:massless} and add some chiral momentum as above%
\begin{equation}
ds^2 = \tilde r^2\left(d\tilde x^+d\tilde x^- + k\,\frac{\ell^2_3}{\tilde r^2}(d\tilde x^-)^2\right) +
\ell^2_3\frac{d\tilde r^2}{\tilde r^2}\,.
 \label{eq:extbtz}
\end{equation}
The latter is equivalent to the extremal massive BTZ black hole
\eqref{eq:btz} (with $\tilde r^2=r^2-r^2_h$).

\paragraph{Summary:}
The near horizon limit \eqref{eq:limit1} of massless BTZ is the null \ads3 orbifold.
The latter corresponds to a thermal state of vanishing temperature in the dual 2d chiral CFT. Thus it describes the
 $p^+=0$ sector of the DLCQ CFT. Exciting the null orbifold by the addition of chiral momentum generates the spacelike
self-dual orbifold, which appears as the near horizon limit of extremal BTZ. This set of relations is summarized in the diagram \ref{dia1}.

\begin{equation}
  \begin{CD}
    \text{Massless BTZ} @>{\text{+ momentum}}>>  \text{Extremal BTZ}\\
       @VV{\operatorname{\text{near horizon}}}V      @VV{\operatorname{\text{near horizon}}}V\\
    \text{Null orbifold}   @>{\text{+ momentum}}>> \text{Self-dual orbifold}
  \end{CD}
\label{dia1}
\end{equation}

Notice this full discussion corresponds to the construction of non-relativistic gravity duals in d=3. The near horizon limit implements the low energy limit involved in the non-relativistic limit. In d=3, the lack of bulk propagating degrees of freedom manifests itself in a set of different global identifications characterising the different chiral momentum sectors of the DLCQ CFT. These different quotients realise the correct symmetries preserved after the non-relativistic limit of the original d=2 CFT. The only non-singular non-relativistic gravity dual corresponds to the sector of large $p^+$, as is customary in gravity/gauge theory dualities and Matrix theory\cite{BFSS,Balasubramanian:1997kd}.\footnote{In higher dimensions, the easiest construction of non-relativistic gravity duals to DLCQ CFTs is also achieved by exciting momentum through the addition of some wave propagating in the bulk, which in $d>3$ is indeed dynamical \cite{Goldberger:2008vg,Maldacena:2008wh}. We would like to emphasize that such construction had already been used in \cite{Cvetic:1998jf,Brecher:2000pa}.}

\subsection{The pinching \ads3 orbifold as the near horizon limit of massless
BTZ}\label{NH-Pinching-orbifold}

The above discussion is coherent and completes the description of chiral excitations in 2d CFTs.
But, it was apparent that when taking the near horizon of the massless BTZ black hole \eqref{eq:massless}, there was an inequivalent near horizon limit that we could have considered
\begin{equation}
r=\epsilon\rho\,,   \quad \quad \hat{x}^\pm = \epsilon\,x^\pm\,.
\label{eq:pinching}
\end{equation}
The resulting geometry, like \emph{all} geometries we are considering, is locally AdS${}_3$
\begin{equation}
  ds^2 = \rho^2d\hat{x}^+\,d\hat{x}^- + \ell_3^2\,\frac{d\rho^2}{\rho^2}\quad \quad \hat{x}^\pm\sim \hat{x}^\pm + 2\pi\epsilon\,.
\end{equation}
We will refer to it as a pinching \ads3 orbifold due to its interpretation as AdS$_3/\mathbb{Z}_{1/\epsilon}$ where the quotient acts on the spacelike S${}^1$ of AdS${}_3$. By construction, its causal boundary is a \emph{pinching cylinder}, $R\times S^1/\mathbb{Z}_{1/\epsilon}$.
This geometry is singular, but it is {\it not} equivalent to the null self-dual orbifold. In the next section, we will interpret it as a low energy limit in which {\it all} excitations in both chiral sectors are decoupled. Thus, different near horizon limits capture very different physics.

\section{Low energy IR limits of 2d non-chiral CFTs}
\label{2d-CFT-section}%

All black hole geometries considered in the previous section are interpreted as thermal states in some dual 2d non-chiral CFT with finite central $c$. This theory is formulated on a cylinder of radius $R$
\be%
ds^2 = -dt^2 + R^2 d\phi^2 = -du \, dv   ~~~;~~~ u = t - R\phi, \ v
= t + R\phi \label{cylinder}
\ee%
where $\phi$ is a circle, i.e. $\phi\sim \phi + 2\pi$. After quantization, the eigenvalues of the momentum operators conjugate to $u$ and $v$, $P^v$ and $P^u$, respectively, are
 \be\label{Pu-Pv}%
 P^{v} =
\left(h+n-\frac{c}{24}\right)\frac{1}{R},\qquad P^{u} =
\left(h-\frac{c}{24}\right)\frac{1}{R}\,, \qquad  n\in \mathbb Z %
\ee%
where $n$ denotes the quantized momentum along the S${}^1$. Assuming unitarity, implies
$h \geq 0$ and $h + n \geq 0$. These are related to the eigenvalues of the standard
operators $L_0,\bar{L_0}$ used in radial quantization on the plane
by $\bar L_0=h+n$ and ${L}_0 = h$. We will assume that the 2d CFT is
non-singular, and therefore, that its spectrum is discrete.

We now interpret the two near horizon limits studied in the previous section.
\begin{itemize}
\item[1.]
The null orbifold has the same boundary structure as the spacelike self-dual orbifold and corresponds to the $r_h\to 0$ limit of the latter. The first feature freezes out the right moving sector to $h=0$ (or $\bar L_0=\frac{c}{24}$). To see this we note that taking the near horizon limit in the dual 2d CFT amounts to taking the mass gap of the right moving sector to infinity  while keeping the gap in the left moving sector finite \cite{Balasubramanian:2009bg}. The second corresponds to setting the temperature to zero, and consequently, the momentum $p^+=0$.

\item[2.] The pinching orbifold corresponds to sending the radius $R$ of the limiting boundary cylinder to zero,
$R\sim \epsilon\to 0$. This generates an infinite gap in both chiral sectors of the initial 2d CFT. The only surviving finite excitations are those corresponding to $h=n=0$. Thus, given a CFT with a fixed central charge $c$, this limit freezes out (decouples) both
left and right moving sectors, leaving us with the Hilbert space:
\be\label{double-decoupled-Hilbert}%
 {\cal H} = \{  |c/24 \rangle_L\otimes |c/24\rangle_R\}\,.%
\ee%
This CFT limit parallels the pinching orbifold introduced in \ref{NH-Pinching-orbifold}.
To see this, relabel $R=\epsilon \ell$ and write the cylinder metric as
\be%
ds^2=-dt^2+\ell^2 d\psi^2\,, \qquad \phi=\frac{\psi}{\epsilon}\,. %
\ee%
We can indeed interpret the latter as the pinching  cylinder $R\times S^1/\mathbb{Z}_{1/\epsilon}$.

\end{itemize}

The above discussion gives evidence for the claim that different near horizon limits describe different low energy limits in the dual 2d CFT. In particular, given a non-singular unitary 2d CFT with finite central charge $c$, the null self-dual orbifold describes the $p^+=0$ sector of the DLCQ of the original CFT whereas the pinching \ads3 orbifold decouples both chiral sectors. The first has a non-trivial Hilbert space, whereas the second contains {\it no dynamics}, but just the degeneracy of the quantum states $L_0=\bar{L}_0=c/24$. This conclusion is generic unless we scale the central charge of the CFT while taking the low energy limit as we will now discuss.

\section{The pinching orbifolds and double scaling limits}

Sending $R\to 0$ or $c\to\infty$ has opposite effects for the spectrum of the 2d CFT.
The first generates an infinite gap, whereas the second reduces it, generating a
continuous spectrum in the limit. It is natural to wonder whether a double scaling
limit $R\to 0,\,c\to\infty,\, cR=\textrm{fixed}$ could keep some nontrivial dynamics
after the low energy limit discussed in the previous section.

To achieve this we should have some non-zero entropy. Recall
that the geometric entropy of classical BTZ black holes, given by the Bekenstein-Hawking area law, is exactly given
by the semi-classical Cardy formula
\be\label{entropy-CFT-gravity}%
S =\frac{2\pi r_+}{4G_3}
= 2\pi\sqrt{\frac{c}{6}\left(L_0-\frac{c}{24}\right)} +
 2\pi\sqrt{\frac{c}{6}\left(\bar{L}_0-\frac{c}{24}\right)}\,,
\ee
where we have used the standard correspondence\footnote{We remark the entropy funcional dependence on
$c(L_0-\frac{c}{24})$ and $\bar c(\bar L_0-\frac{\bar c}{24})$ products goes beyond lowest order and is valid up
to exponentially suppressed contributions  in the saddle point approximation \cite{Beyond-Cardy}.}
\be\label{L0-barL0}%
L_0-\frac{c}{24}=\frac{M\ell_3+J}{2}=\frac{c}{24}\left(\frac{r_++r_-}{\ell_3}\right)^2\,,\quad
\bar L_0-\frac{c}{24}=\frac{M\ell_3-J}{2}=\frac{c}{24}\left(\frac{r_+-r_-}{\ell_3}\right)^2\,.
\ee%

The first equality in \eqref{entropy-CFT-gravity} suggests that if together with taking a nearly-massless BTZ, i.e. $r_\pm\to \epsilon r_\pm $ with $\epsilon\to 0$, we also scale $G_3\to \epsilon G_3$, the entropy does not change and {\it remains finite}.
The second equality suggests the scaling $c\to c/\epsilon$ (where we used the Brown-Henneaux formula $c=3\ell_3/(2G_3)$ \cite{brown-henneaux}), and then \eqref{L0-barL0} imply $L_0-\frac{c}{24}\to \epsilon(L_0-\frac{c}{24})$ and $\bar{L}_0-\frac{c}{24}\to \epsilon(\bar{L}_0-\frac{c}{24})$.

In what follows we first discuss the geometric argument in terms of BTZ metrics to connect the above scalings to the
pinching orbifold. Afterwards we discuss the meaning of the limit in
the dual field theory, and finally we mention some subtleties involved in the latter.

\subsection{The gravity perspective}\label{pinching-resolution-gravity}

Before introducing the actual double scaling limit in gravity, let us review an interesting observation. Consider the BTZ metric
\be\label{BTZ-l-units}%
ds^2=-F({r})dt^2
+\frac{dr^2}{F(r)}+r^2\left(d\phi-\frac{r_+r_-}{r^2}\frac{dt}{\ell_3}\right)^2\,,
\ee%
where
$$F(r)=\frac{(r^2-r^2_+)(r^2-r^2_-)}{\ell_3^2 r^2}\quad \text{and}\quad \phi\sim \phi + 2\pi\,.$$
The $\lambda$-{transformation}%
\be\label{lambda-scaling}%
r=\lambda \rho,\quad r_{\pm}=\lambda\rho_{\pm},\quad t=\lambda^{-1}\tau,\quad
\phi=\lambda^{-1}\psi \,,%
\ee%
formally keeps the form of the metric invariant as in
\eqref{BTZ-l-units}, but the periodicity in $\psi$ is
$2\pi\lambda$. Note this was not simply a coordinate transformation, as the parameters $r_\pm$
also changed. Denoting a BTZ black hole with charges $\{M,\,J\}$ and periodicity $2\pi\alpha$ by
BTZ$(M,J;2\pi\alpha)$, we learn that in classical gravity
\be\label{BTZ-orbifolding}%
\textrm{BTZ}(M\lambda^2,J\lambda^2;2\pi)\equiv\textrm{BTZ}(M,
J;2\pi\lambda)\,, %
\ee%
where we used \eqref{mass-J}. Notice the {\it entropy} remains {\it invariant} under this $\lambda$-transformation whereas the Hawking temperature gets scaled by $\lambda^{-1}$ \cf \eqref{TL-TR}. Both solutions in \eqref{BTZ-orbifolding} belong to the
{\it same gravity theory} since both $\{\ell_3,\,G_3\}$ remained unchanged. Thus, the central charge did not transform and consequently, we did not yet achieve the goal set at the beginning of this section.

\paragraph{The double scaling limit :} If we want to be left with a finite entropy after taking the near horizon of a nearly massless BTZ black hole, all we have to do is to combine the near horizon limit described previously with an scaling of the 3d Newton's constant as
\be\label{NH-near-massless}%
r_\pm=\epsilon \rho_\pm\,,\qquad r=\epsilon\rho_++\epsilon\rho\,,\qquad t=\epsilon^{-1}\tau\,, \qquad   \phi=\epsilon^{-1}\psi\,,\qquad G_3=\epsilon \tilde G_3 \qquad
\epsilon\to 0\,.
\ee%
This is nothing but a $\lambda$-transformation \eqref{lambda-scaling} with $\lambda=\epsilon$, accompanied by the Newton's constant extra scaling. The latter is the responsible to keep the entropy of the nearly massless BTZ finite.
As shown in the appendix, the corresponding near horizon geometry, which is insensitive to the rescaling of $G_3$, is a {\it pinching orbifold}. The scaling in Newton's constant does scale the central charge as we wanted, i.e. $c\to c/\epsilon$, at the price of changing the actual conformal field theory description of the system, as we discuss next.

\subsection{The 2d CFT perspective}\label{pinching-resolution-CFT}

As discussed the equivalent of the scaling \eqref{NH-near-massless} in terms of 2d CFT quantities is
to rescale the central charge by a factor of $\epsilon^{-1}\equiv K$,
and $L_0-c/24$ and $\bar{L}_0-c/24$ by a factor of $\epsilon=K^{-1}$. As this scaling changes the central charge  it takes us to a different 2d CFT.  One can gain some intuition  about these different 2d CFT's by thinking about the CFT dual to the D1-D5 system. This 2d CFT can be described
by a 2d sigma model with $N=(4,4)$ supersymmetry with a target which can be thought of as a suitable
symmetric product ${\rm Sym}^{N_1N_5}({\cal M}_4)$. Rescaling the central charge is like rescaling
$N_1N_5$. Therefore, we expect a relation of the form
\be \label{j11}
{\rm CFT}_{\rm new} \approx {\rm Sym}^{K}({\rm CFT}_{\rm old}).
\ee
This clearly only makes sense when $K=1/\epsilon$ is an integer, and in that case the new CFT
has a long string sector which is directly inherited from the old theory. The Virasoro
algebra related to this long string sector can be directly related to that of the original
CFT using standard orbifold technology. Explicitly,
given the Virasoro algebra
\begin{equation}
  [L_n,\,L_m] = (n-m)L_{n+m} + \frac{c}{12}n(n^2-1)\delta_{n+m}
\end{equation}
consider the subalgebra with generators\footnote{Notice that the apparently strange transformation for the generator $l_0$ is due to the fact that we were working on the plane. Indeed, if we would have worked on the cylinder, the transformation is the expected one :
$$
  l^{\text{cyl}}_n \equiv \frac{1}{K} L^{\text{cyl}}_{nK}\,, \quad n\neq 0\,, \quad \quad
  l^{\text{cyl}}_0 \equiv \frac{1}{K} L^{\text{cyl}}_{0}\,.
$$
We now see that the transformation quoted on the plane makes sure the above cylinder transformation brings us back to the plane.}
\begin{equation}
  l_n \equiv \frac{1}{K}\,L_{nK}\,, \quad n\neq 0\,, \quad \quad \quad l_0 \equiv \frac{1}{K}(L_0-\frac{c}{24}) + \frac{c}{24} K\,.
\end{equation}
It is then straightforward to see that $l_n$ also form a Virasoro algebra with central charge $c^\prime = cK$ and
that the spectrum of $l_0$ has a spacing of $1/K$ compared to that of $L_0$.

Since the long string sector in an orbifold theory tends to dominate the entropy, the equivalence of the long string sector in the orbifold theory to the original CFT provides a natural explanation of the constancy of the entropy.

\subsection{Some caveat comments}

The previous discussion was all at the level of supergravity, and one may wonder how
much of this survives when considering the full quantum theory. For special CFT's and
special values of $c,L_0$ and $\bar{L}_0$ a much more precise statement can be made.
For example, for the extremal BTZ black hole in the D1-D5 system, with $L_0-\bar{L}_0=N_p$,
there are U-dualities which replace $N_1N_5$ by $KN_1N_5$ and $N_p$ by $N_p/K$, as long
as $N_1N_5$ and $N_p$ are coprime and $K$ divides $N_p$. These U-dualities do not
exist for arbitrarily large values of $K$ since $K$ cannot exceed $N_p$. To be able
to take $K$ to infinity we need to consider non-extremal and non-supersymmetric
black holes and consider excitations with $L_0=\bar{L}_0\neq c/24$,
since $L_0-\bar{L}_0$ is quantized,
but no suitable dualities that do the job for these systems are known.

It is also important to keep in mind that once $L_0,\bar{L}_0$ become comparable to $c/24$,
the original BTZ metric is presumably no longer a good description of the system.
For these values of the parameters there can be Gregory-Laflamme instabilities
to localisation on some of the internal compact directions that we suppressed in our
discussion\footnote{For an explicit example, see \cite{deBoer:2008fk}.}, and in addition
higher order terms in the space-time effective action
may become important. It is therefore not at all clear that the $\epsilon\to 0$
limit we took will make sense in the full quantum theory.

\section{Discussion and outlook}

In this paper we have discussed two possible near horizon, low
energy, limits of  nearly massless BTZ black holes. If massless BTZ
is viewed as a special case of extremal BTZ, in the near horizon
limit one obtains a null self-dual \ads3 orbifold which is a special case of the
spacelike self-dual orbifold with zero  momentum. In the dual 2d CFT, and in
light of the discussions in \cite{Balasubramanian:2009bg}, this means
that the null orbifold is dual to the zero light-cone momentum
$p^+=0$ sector of the DLCQ of the CFT. Whether the singular
and ill-defined null orbifold metric will eventually shed
any new light on the equally
singular and ill-defined
$p^+=0$ sector/state remains to be seen. Our point here was to
establish the connection between the two. The second limit gave rise to the
singular pinching \ads3 orbifold, whose dual 2d CFT interpretation corresponds to a
decoupling of both chiral sectors of the theory leaving us with the states $L_0=\bar{L}_0=c/24$.
We comment that, as discussed in the appendix, the difference between the two cases can also be  viewed as an order of limits in taking the near horizon limit of near extremal BTZ black hole. To see this, start with a generic BTZ and scale $r_+$ and $\delta\equiv r_+- r_-$ to zero. One may first scale $\delta$ to zero and then take $r_+$ to zero (that is scale $\delta$ to zero faster than $r_+$) or scale them to zero at the same rate.  The former leads to the null self-dual orbifold while the latter to the pinching AdS$_{3}$ orbifold.

Another point we did not address so far is the fact that
the $M=0$ BTZ black hole does not describe the true ground
state of the system, which is global \ads3. One may wonder
whether there is also a geometry which describes the ``near-horizon"
limit of geometries with $L_0<c/24$, i.e. conical defect metrics.
The natural candidate is to take (\ref{eq:c1}) and to continue
$r_h^2$ to negative values. This will give rise to timelike orbifolds
with closed timelike curves, which are usually linked to non-unitarity of the dual CFT.
We would normally discard such geometries as unphysical and leave a further
understanding of these to future work.

We discussed that the nearly massless BTZ black hole can also be
viewed as a BTZ black hole with parametrically finite mass and
angular momentum over a pinching \ads3 orbifold. The natural units
to measure the inner and outer horizon radii $r_\pm$ of a BTZ black
hole are the \ads3 radius units $\ell_3$. Therefore, it is
intuitively expected that the smallness of the radii could be
resolved if one used a different set of units. In our case 3d Newton
constant $G_3$ is the other natural scale in the problem. In this
units and in a suitable $G_3\to 0$ limit, such that $r_\pm/G_3$
remains finite, radii will remain finite. We used this to argue the existence of a
double scaling limit under which  non-trivial degrees of freedom survive.

\paragraph{Non-extremal excitations and Rindler physics :} It is natural to consider whether similar considerations would provide new insights for non-extremal BTZ black holes. In this case, there may also be different near horizon limits that one could take, but if we consider
\begin{equation}
  r = r_+ + \epsilon\,\frac{1}{2}\frac{r_+^2-r_-^2}{\ell_3^2\,r_+}\,y^2\,, \quad \frac{t}{\ell_3} = \frac{r_+\,\ell_3}{r_+^2-r_-^2}\,\tau\,, \quad \phi = \frac{r_-\,\ell_3}{r_+^2-r_-^2}\,\tau + \frac{\sqrt{\epsilon}}{r_+}\,\psi \quad \quad\epsilon\to 0
\end{equation}
the resulting metric is :
\begin{equation}
  ds^2 \to \epsilon\,\left(y^2\,d\tau^2 + dy^2 + d\psi^2\right) \quad \quad \quad \psi\sim \psi + 2\pi\,\frac{r_+}{\sqrt{\epsilon}}\,.
\end{equation}
The overall $\epsilon$ scaling can be interpreted as a limit where $\ell_p\to
0$, i.e. $\ell_p=\sqrt{\epsilon}\,\ell_p^\prime$. Notice that in
Planck units, the periodicity of the angular variable $\psi$ is {\it
finite}. Thus, the near horizon limit is two dimensional Rindler
space times an S${}^1$. Applying the Bekenstein-Hawking formula to
the area of the Rindler horizon at $y=0$, one is able to reproduce
the entropy of the original non-extremal black hole.

The rescaling of the 3d Planck length, or equivalently, the 3d Newton
constant, can perhaps once more be interpreted as passing to the
long string sector of an orbifold theory, and it would be very
interesting if this could be used to shed further light on a possible
holographic dual description of Rindler space  and hence a generic non-extremal black hole.

\paragraph{Relevance for the microscopics of extremal black holes :} One of our motivations to consider the vanishing mass BTZ black hole was as a prototype of an extremal vanishing horizon black hole (EVH). Given the difficulties to prove/disprove the Kerr/CFT proposal, it is natural to move in the moduli space of extremal solutions to find points where local AdS${}_3$ throats appear. It was already noted in \cite{Bardeen-Horowitz} that such (singular) throats exist for d=5 extremal Kerr. In fact, the singular geometry reported there is an example of the embedding of our pinching orbifold in d=5. More examples of such near extremal vanishing black holes in AdS${}_5$ and AdS${}_4$ backgrounds were studied in \cite{FGMS, Balasubramanian:2007bs}. More recently, they have also appeared in \cite{compere,guica,terashima,Matsuo-Nishikoa}. The
analysis and conclusions reported here are therefore relevant to understand the different physics involved in the different near horizon geometries emerging in these singular situations.
We will take first steps in this direction in an upcoming publication \cite{EVH-us}.


\section*{Acknowledgements}
We would like to thank Arjun Bagchi, Vijay Balasubramanian, Geoffrey Compere and James Lucietti for useful discussions. J.S. would like to thank Simon Ross and Jose M. Figueroa-O'Farrill for discussions on AdS quotients many years ago. The work of J.S. was partially supported by the Engineering and
Physical Sciences Research Council [grant number EP/G007985/1]. The work of J.d.B. was partly supported by the Foundation of Fundamental Research on Matter (FOM).

\appendix

\section{Near-extremal vanishing horizon limits for BTZ black holes}\label{app:BTZ}

In this appendix, we discuss the geometries obtained as near horizon limits of (near-)extremal vanishing horizon BTZ black holes as a function of the speed at which the horizon and the non-extremality vanish. As discussed in the main text for massless BTZ, we already know of the existence of two different ways of taking these limits : one giving rise to the null self-dual orbifold and a second one giving rise to the pinching orbifold. We will show that there exist pinching orbifold versions for all vanishing horizon limits in AdS${}_3$. Interestingly, we show that due to the self-dual nature of the null orbifold, the latter does not extend to arbitrary near-extremal vanishing horizon limits.

\subsection{Pinching orbifolds as near horizons of near-extremal vanishing horizon BTZ}
\label{appA}

We discuss the different extremal vanishing horizon limits for the family of BTZ black holes
\begin{equation}
  ds^2(L)= - \frac{(r^2-r_+^2)(r^2-r_-^2)}{L^2r^2}dt^2 + \frac{L^2r^2}{(r^2-r_+^2)(r^2-r_-^2)}dr^2 + r^2\left(d\phi - \frac{r_+r_-}{Lr^2}dt\right)^2
\label{eq:btz2}
\end{equation}
in an asymptotically $AdS_3$ spacetime of radius of curvature $L$.

To discuss the different extremal vanishing horizon limits, we introduce the notation
\begin{equation}
  r_\pm^2 = r_\star^2 \pm \delta r_\star^2 \quad \quad r_\star = \epsilon^\alpha\,\rho_\star\,, \quad \delta r_\star = \epsilon^\gamma\,\delta\rho_\star
\end{equation}
with $\gamma\geq \alpha$. Thus, $\alpha$ and $\gamma$ control how fast the horizon and non-extremality vanish, respectively. We define the near-horizon limit as
\begin{eqnarray}
  r &=& r_\star + \epsilon^\beta\,y \equiv \sigma\,\epsilon^{\text{min}(\alpha,\beta)}\,, \\
  r + r_\star &=& 2r_\star + \epsilon^\beta\,y \equiv \hat{\sigma}\,\epsilon^{\text{min}(\alpha,\beta)}
  \,, \\
  t &=& \epsilon^{-\beta}\,\tau \,, \\
  \phi &=& \epsilon^{-\text{min}(\alpha,\beta)}\,\psi + \frac{r_+r_-}{Lr_\star^2}t\,, \quad \quad \epsilon\to 0\,.
\end{eqnarray}
The resulting finite metric is
\begin{equation}
  ds_{\text{nh}}^2(L) = -\frac{\left(y^2\hat{\sigma}^2-\epsilon^{2\kappa}\delta\rho_\star^4\right)}{L^2\sigma^2}\,d\tau^2 + \frac{L^2\sigma^2}{\left(y^2\hat{\sigma}^2-\epsilon^{2\kappa}\delta\rho_\star^4\right)} + \sigma^2\,\left(d\psi + \frac{y\hat{\sigma}}{L\sigma^2}{\cal B}\,d\tau\right)^2
 \label{eq:nhevh}
\end{equation}
where
\begin{equation}
  {\cal B} = \sqrt{1-\epsilon^{4(\gamma-\alpha)}\frac{\delta\rho_\star^4}{\rho_\star^4}}\,, \quad \quad \kappa = 2\gamma - \beta - \text{min}(\alpha,\,\beta)\,.
\end{equation}
What is important to stress is that if the horizon vanishes $(\alpha > 0)$ and we consider a near horizon limit $(\beta>0)$, the resulting metric is an orbifold of AdS${}_3$, and it will always be a pinched version of a known quotient given the $\epsilon$ dependent periodicity
$\psi \sim \psi + 2\pi\,\epsilon^{\text{min}(\alpha,\,\beta)}$.

When $\gamma\geq \alpha> 0,\  \kappa,\ \beta\geq 0$, one distinguishes different possibilities:
\begin{itemize}
\item[1.] {\it Pinching Self-dual orbifold} corresponds to $\beta>\alpha>0,\ \kappa>0$, i.e. $\gamma>\alpha$. Thus,
\begin{equation*}
\sigma= \frac{\hat\sigma}{2}= \rho_\ast\,,\quad {\cal B}=1\,,%
\end{equation*}
and the near-horizon metric is
\begin{equation}\label{alpha<beta}%
 ds^2(L) = -4\frac{y^2}{L^2} d\tau^2+\frac{L^2}{4} \frac{dy^2}{y^2}+\rho_\ast^2 (d\psi
+ 2\frac{y}{L\rho_\ast} d\tau)^2 \,.
\end{equation}
The three dimensional part of \eqref{alpha<beta} is a spacelike self-dual
AdS$_3$ orbifold, except for the fact that  $\psi$ has periodicity
$2\pi \epsilon^\alpha$.

\item[2.] {\it Pinching near self-dual orbifold} corresponds to $\beta>\alpha>0,\ \kappa=0$. In
this case  $2\gamma=\alpha+\beta$ and the metric has the same form
as \eqref{alpha<beta} except for the fact that $y^2$ is replaced by
$y^2-\frac{\delta\rho_\ast^2}{4\rho_\ast^2}$ in the two dimensional base spanned by $\{\tau,\,y\}$. These are the pinching analogues of the so-called AdS${}_2$ black holes, for a recent discussion of the latter see
\cite{finn1,finn2}.

\item[3.] {\it Pinching Massless BTZ} corresponds to $\alpha>\beta>0$. In this case
$\kappa>0$ and
\begin{equation*}
\sigma=\hat\sigma=y,\quad {\cal B}=1\ ,%
\end{equation*}
The near-horizon    metric becomes
\begin{equation}\label{alpha>beta}%
\begin{split}
ds^2(L) &= -\frac{y^2}{L^2} d\tau^2+L^2\frac{dy^2}{y^2}+ y^2 (d\psi
+ d\tau/L)^2\\
&=L^2\frac{dy^2}{y^2}-y^2 dx^+ dx^-\,,
\end{split}
\end{equation}
where $x^-=\psi$ and $x^+=2\tau/L+\psi$.
This metric is that of a massless BTZ black hole \eqref{eq:massless}, except for the fact
that now $x^-\sim x^-+ 2\pi \epsilon^\beta$.

\item[4.] {\it Pinching Extremal BTZ} corresponds to $\alpha=\beta>0,\ \kappa>0$. In
this case $\gamma>\alpha$ and
\begin{equation*}
\sigma=\rho_\ast+y,\quad \hat\sigma=2\rho_\ast+y,\quad {\cal B}=1\,.%
\end{equation*}
The near-horizon    metric is
\begin{equation}\label{alpha>beta2}%
ds^2(L) = - \frac{(\sigma^2-\rho_\ast^2)^2}{\sigma^2 L^2} d\tau^2+L^2\frac{\sigma^2d\sigma^2}{(\sigma^2-\rho_\ast^2)^2}+
\sigma^2 (d\psi + \frac{\sigma^2-\rho_\ast^2}{\sigma^2}d\tau)^2
\end{equation}

\item[5.] {\it Pinching Non-extremal BTZ} corresponds to $\alpha=\beta>0,\ \kappa=0$. In
this case $\gamma=\alpha$ and
\begin{equation*}
\sigma=\rho_\ast+y,\quad \hat\sigma=2\rho_\ast+y,\quad {\cal B}=
\left(1-\frac{\delta\rho_\ast^4}{\rho_\ast^4}\right)^{1/2}\ ,%
\end{equation*}
for which the near-horizon    metric reduces to%
\begin{equation}\label{alpha=beta=gamma}%
 ds^2 = -h(\sigma^2)
 d\tau^2 +  \frac{d\sigma^2}{h(\sigma^2)} +
\sigma^2 (d\psi + {\cal B}
\frac{{\sigma}^2-\rho_\ast^2}{\sigma^2} d\tau)^2
\end{equation}
where
$$
h(\sigma^2)=\frac{(\sigma^2-\rho_\ast^2-\delta\rho_\ast^2)
(\sigma^2-\rho_\ast^2+\delta\rho_\ast^2)}{\sigma^2 L^2}\ .
$$
\end{itemize}
Thus, for all allowed near-extremal vanishing horizon BTZ black holes, there exists a pinched orbifold near horizon geometry.

\subsection{Null orbifold limits}

Let us consider an extremal BTZ black hole
\begin{equation}
  ds^2 = -\frac{(r^2-r_+^2)^2}{r^2\,L^2}\,dt^2 + \frac{L^2\,r^2}{(r^2-r_+^2)^2}\,dr^2 + r^2\left(d\phi - \frac{r_+^2}{r^2}\,\frac{dt}{L}\right)^2\,,
\end{equation}
with horizon scaling to zero as $r_+=\epsilon^\alpha\,\rho_+$ and study the near horizon limit
\begin{equation}
  r^2 = \epsilon^{2\alpha}\rho_+^2 + \epsilon\,\rho\,, \quad \quad \epsilon\to 0\,.
\label{eq:ghor}
\end{equation}
\begin{itemize}
\item[1.] If $2\alpha > 1$, this reduces to the massless BTZ calculation discussed in the main text. Thus, we do get a null-self-dual orbifold if we scale one of the lightline coordinates.
\item[2.] If $2\alpha < 1$, we complete the near horizon limit \eqref{eq:ghor} with
\begin{equation}
  \phi = \varphi + \frac{\tau}{\epsilon\,L}\,, \quad \quad t=\frac{\tau}{\epsilon}\,.
\end{equation}
As for the vanishing horizon limit of the spacelike self-dual orbifold discussed in the main text, the quadratic piece in $d\varphi^2$ vanishes in the limit, $dt^2$ pieces cancel and we are only left with the cross-term $d\varphi\,d\tau$, which is the signature of the null self-dual orbifold, since $\varphi\sim \varphi + 2\pi$.
\item[3.] If $2\alpha=1$, we also complete the near horizon limit \eqref{eq:ghor} with
\begin{equation}
  \phi = \varphi + \frac{\tau}{\epsilon\,L}\,, \quad \quad t=\frac{\tau}{\epsilon}\,.
\end{equation}
In this case, it is natural to work wit the new radial coordinate $\sigma=\rho_+ + \rho$. The resulting metric is a null self-dual orbifold
\begin{equation}
  ds^2 = \frac{L^2}{4(\sigma-\rho_+^2)^2}\,d\sigma^2 + 2(\sigma-\rho_+^2)\,d\varphi\,\frac{d\tau}{L}\,.
\end{equation}
\end{itemize}

It is interesting to study whether this null self-dual orbifold extends to {\it near-extremal} vanishing horizon limits. This point will be further discussed in \cite{EVH-us}. Here, we will just consider the particular near horizon, vanishing horizon, scalings
\begin{equation}
r^2=\epsilon \rho\,,\quad r^2_\pm=\epsilon\rho_\pm\,,\quad \epsilon\to 0\,.
\end{equation}
It can be shown that whenever $\rho_+\neq \rho_-$, it is {\it not} possible to find a finite non-degenerate metric by scaling time $t$ and shifting the angular variable $\phi$ while keeping its periodicity independent of $\epsilon$ to avoid the generation of any of the pinching orbifolds discussed previously. In other words, if the amount of non-extremality is large enough, the null self-dual structure emerging in the near horizon disappears.


\providecommand{\href}[2]{#2}\begingroup\raggedright

\endgroup

\end{document}